\documentclass[11pt,twoside]{article}

\usepackage{asp2006}

\begin{document}
\title{On the role of relativistic effects in X-ray variability of AGN}   
\author{Piotr T. \.{Z}ycki}   
\affil{Nicolaus Copernicus Astronomical Center, Warsaw, Poland}    
\author{Andrzej Nied\'{z}wiecki}   
\affil{{\L}\'{o}d\'{z} University, Department of Physics, {\L}\'{o}d\'{z}, 
 Poland}    

\begin{abstract} 
Special relativistic and strong gravity effects are clearly seen in X-ray 
energy 
spectra from AGN. Most important here are the broad profiles of the Fe
K$\alpha$ line observed in a large fraction of sources. These indicate
that X-ray generation and reprocessing takes place very close to the
central black hole. Here we explore
consequences of such effects on X-ray variability. We perform computations of a
possible quasi-periodic signal from a Keplerian motion of primary X-ray source.
We also study in some details the light-bending model of variability of 
the X-ray reprocessed component, extending previous work on the subject.

\end{abstract}


\section{Introduction}   

X-ray emission from AGN is thought to be produced in the very central regions
of accretion flow, close to the black hole, since this is where most of the 
gravitational potential energy is dissipated. Observational evidences for
this come from results of modeling of  spectral features, often showing 
distortions consistent with those caused by Doppler effects and gravitational
redshift. In particular, broad profiles of the Fe K$\alpha$ line are currently
the best evidences for the very central X-ray generation and reprocessing.

Geometrical models of X-ray emitting accretion flows are of two kinds:
the accretion disk with a corona scenario, and a hot flow scenario. These were
extensively studied in the context of X-ray binaries, where the wealth of 
data enables putting stronger constraints on the geometry of accretion flow
than in the case of AGN. In the former scenario the standard Keplerian 
accretion disk extends to the last stable orbit and the hard X-rays are 
produced by an active corona above the disk. This is motivated mainly by the
solar corona where magnetic flares are thought to be powered by differential
rotation of the Sun's interior. The hot flow scenario is based on an 
alternative solutions of accretion flow, where the flow is optically thin 
and hot, 
most likely two-temperature, with proton temperature reaching the virial value.
Both the observational data and the theoretical arguments suggest that the 
two solutions operate in different ranges of mass accretion rate.

Presence of spectral features from X-ray reprocessing by cold optically
thick plasma, with strong relativistic distortions obviously implies 
that the disk has to extend to the last stable orbit. This is the situation
in, e.g., the Seyfert 1 galaxy MCG--6-30-15, which is currently the best 
example of effects of strong gravity in X-ray spectra. In this contribution
we study some of the consequences of the relativistic effects on the
variability of X-ray emission.

\section{Quasi-periodic modulation of X-rays from rotating sources}
\label{zyckisec:qpo}

Assume that the hard X-rays are produced by magnetic flares co-rotating
with the Keplerian disk. Assume further that durations of the flares
are consistent with the flare avalanche model of \citet{pf99},
so that there is a broad distribution of the durations (superposition of
flare profiles of different durations explains the slope of the power 
spectrum).
Then, the natural association of a flare time scale with the Keplerian
time scale (so that longer flares are produced further away), results 
in the radial X-ray emissivity increasing with radius, which is clearly 
inconsistent with the dissipation rate 
of gravitational energy \citep{ptz02}. Abandoning this association 
leads to a possibility of long flares occurring at small distances 
($t_{\rm flare} \gg t_{\rm Kepler}$). The X-rays source can then participate
in the Keplerian motion, possibly for a number of $t_{\rm Kepler}$, which
produces a periodic modulation of emission due to Doppler boost.
The strength of the resulting quasi-periodic signal in the PDS depends
on the black hole spin, inclination angle and other details of the 
radial distribution of flares. For parameters appropriate for the Seyfert
galaxy MCG--6-30-15, where $M\approx 10^6\,{\rm M}_\odot$ and $i\approx 
30^\circ$,
the strength of the predicted signal is high enough for the signal,
if present, to be detected in the {\it XMM-Newton\/} data \citep{ptzan05}. 
However, the data \citep*{vaug03}
do {\em not\/} reveal such a signal. This would suggest that the X-rays
cannot be produced by co-rotating flares, but rather by a structure
not participating in the Keplerian rotation, e.g., the base of a jet.
On the other hand, a number of observations show periodic variability of 
the energy of a narrow Fe K$\alpha$ line, suggesting line production in 
localized regions co-rotating with the disk (T. J. Turner, these proceedings).
If the two properties -- no QPO signal in the primary emission and
a periodic modulation of the Fe line -- turn out to be universal, this
will clearly suggest a non-trivial geometry of the X-ray source.

\section{The light-bending model of variability of the reprocessed
component}

The reprocessed component (Fe K$\alpha$ line and the Compton reflected
continuum) show rather weaker variability than, and uncorrelated with,
the variability of the primary emission \citep[e.g.,][]{vf04}. 
This is observed on a time scale of $10^4$--$10^5$ sec, i.e., {\em longer\/}
by a factor of $\approx 10$--100 than the Keplerian timescale at the inner 
disk (assuming $M=10^6\,{\rm M}_\odot$). 
Such a reduction of variability is rather surprising, since in the 
simplest model the reprocessed component  should
respond to primary variations  on the light travel timescale. The latter
is very short, if indeed the reprocessing takes place close to the center,
as evidenced by the broad profiles of the Fe K$\alpha$ line.

The light bending model, formulated by \citet[hereafter MF]{mf04},
postulates that the de-coupling of variations is due to changing amount
of relativistic effects, in particular bending of light trajectories from 
the source
to the accretion disk, as the location of the source changes. In its basic
form the model assumes that the intrinsic emission is constant, and 
the observed variability is solely due to changing location of the primary
source. The variability of the reprocessed component is reduced because
reprocessing takes place in a more extended area. \citeauthor{mf04}
demonstrate that in the simple geometry of an off-axis source moving
vertically, the model predicts reduction of line variability of a factor
of $\approx 7$, as the source height changes from 1 to $20 R_{\rm g}$.

\begin{figure}[!ht]\
\plotfiddle{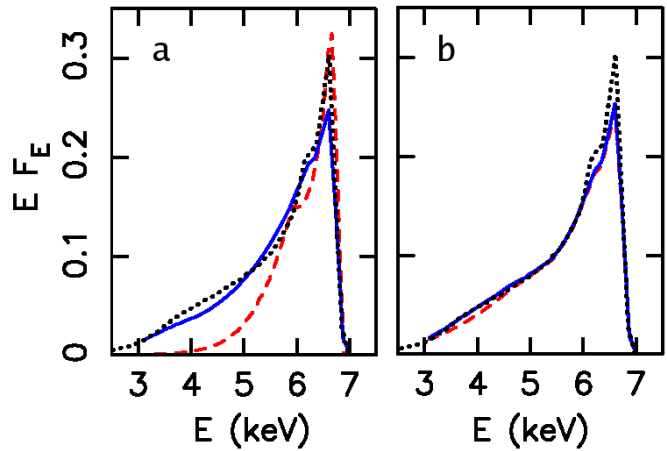}{6cm}{0}{33}{33}{-160}{-20}
\caption{Observed profiles of the Fe K$\alpha$ line with extended red wing 
require an X-ray source located at very low height above the disk.
This is shown in (a) for model C -- an off axis source moving vertically, 
and in (b) for model S -- a source moving radially. The dotted (black) curve 
shows the best fit profile to MCG--6-30-15 data. 
In (a) the solid (blue) curve is a superposition of profiles
for $h_{\rm s}=0.2$--$0.6\,R_{\rm g}$, while the dashed (red) profile is
for $h_{\rm s}=8\,R_{\rm g}$ (as in MF). In (b) the solid (blue)profile
is for $h_{\rm s}=0.07 r$ and $r=2$--$3\,R_{\rm g}$.
\label{zyckifig:kaprof}}
\end{figure}

We have extended the MF model to consider a number of geometrical scenarios:
in addition to the basic model (C -- the same as in MF), where the source 
is located off-axis, at projected radial distance $\rho_{\rm s}$, and
moves vertically, we considered an on-axis source (model A) moving vertically
and a source located very low above the disk surface, moving {\em radially\/}
(model S). Photon transfer in the Kerr metric was computed with the code
described in detail in \citet{ptzan05}.
We first computed time averaged profile of the Fe K$\alpha$ line in order 
to find the range of parameters where the profile match that observed in 
MCG--6-30-15. Then, we computed variability of the line vs.\ that of the 
primary emission, as the source location changes. Details are presented
in Nied\'{z}wiecki \& \.{Z}ycki (in preparation).

\begin{figure}[!ht]
\plotfiddle{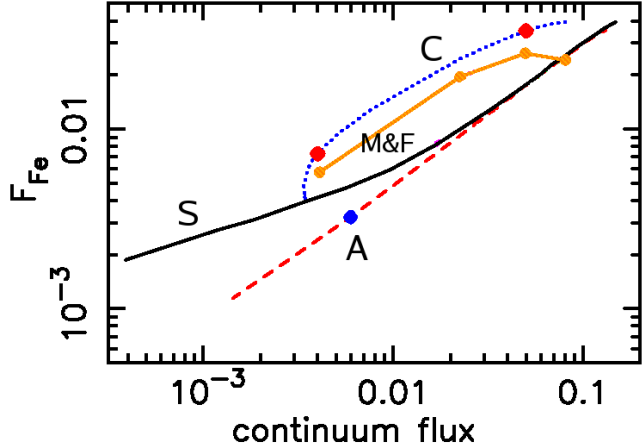}{6cm}{0}{90}{90}{-150}{-10}
\caption{The Fe K$\alpha$ line vs.\ continuum flux diagram showing changes
of the flux as the position of an X-ray source changes. Model S predicts
most significant reduction of the line variability. Original computations
of Miniutti \& Fabian (model C) overestimate the reduction effect.
\label{zyckifig:ff}}
\end{figure}

Our results are somewhat different than those of MF. For $a=0.998$ and 
$i=30^\circ$, in model C (with $\rho_{\rm s}=2\,R_{\rm g}$), we find
that the observed line profile requires a very low height of the source,
$h_{\rm s} \la 1\,R_{\rm g}$. The profile computed for
$h_{\rm s} = 8\,R_{\rm g}$ (as in MF) has a rather weak red wing, certainly
not compatible with what is observed in MCG--6-30-15 
(see Fig.~\ref{zyckifig:kaprof}a). Similar profiles can be obtained in 
model S, where the source {\em radial\/} position changes around 
2--3$R_{\rm g}$ (Fig~\ref{zyckifig:kaprof}b). 

We then addressed the problem of reduced variability of the K$\alpha$ line
and we computed the luminosity of the primary source vs.\ that of the Fe 
K$\alpha$ line, as the source position is changed. Our results are plotted
in Fig.~\ref{zyckifig:ff} for a number of models. In model C we find
no reduction of variability of the line in contrast to results of MF,
especially for source height $h_{\rm s}\sim 10\,R_{\rm g}$.
This is most likely caused by differences in the assumed value of the outer
disk radius: $600\,R_{\rm g}$ in our case compared to $100\,R_{\rm g}$ in 
MF computations. This latter value is too small for the line to be
computed correctly when the height of the source is $\ga 10 \, R_{\rm g}$.

We find a significant reduction of line variability in model S, for 
a source moving radially very low above the disk surface. Source position
changing between $r_{\rm s}=1.23$--$4\,R_{\rm g}$ produces line flux changes 
by a factor of $\approx 7$ compared to a factor of $\approx 100$ for the
continuum. This is primarily a result of light bending to the equatorial
plane, beyond $r \approx 6$, where a relatively constant blue horn
is then produced (Nied\'{z}wiecki \& \.{Z}ycki, in preparation). 
Note, though, that model S would predict the strong
signal in PDS (on Keplerian timescale) which is not observed 
(Sec~\ref{zyckisec:qpo}).


\acknowledgements 

This work  was partly supported by grant no.\ 2P03D01225
from the Polish Ministry of Science and Higher Education.


\end{document}